\documentclass[preprint,showpacs,preprintnumbers,amsmath,amssymb]{revtex4}

\usepackage{graphicx}
\usepackage{dcolumn}
\usepackage{bm}

\begin{document}

\title{Analyses of reflection and transmission at moving potential step}

\author{Toshiharu SAMURA}
\email{samura@akashi.ac.jp}
\author{Masato OHMUKAI}
\email{ohmukai@akashi.ac.jp}

\affiliation{%
Department of Electrical and Computer Engineering, Akashi National
College of Technology, 679-3, Nishioka, Uozumi-cho, Akashi, Hyogo,
674-8501, Japan
}%


\date{\today}

\begin{abstract}
The reflection and transmission of wave functions at a potential step is
 a well-known issue  in a textbook of quantum mechanics. We studied 
 the reflection and transmission characteristics analytically when the potential
 step is moving at a constant velocity $v$ in the same direction as an
 incident wave function by means of solving the time-dependent
 Schr\"{o}dinger equation. As for an infinite potential step, it is known
 that group velocity is
 the same as the moving velocity of the potential step. We found two
 interesting results when the potential step has a finite height of $V_0$. The transmission
 occurs when the kinetic energy of incident wave function is larger than
 the effective potential hight of 
 $\left( \sqrt{\frac{m}{2}}v + \sqrt{V_0}\right)^2$.  The other result is that the reflectivity
 depends on
 $x$, which derives from the interference between the incident and the
 reflected wave functions.

\end{abstract}

\maketitle

\section{Introduction}

The reflection and transmission of wave functions at a potential step is
one of the most
fundamental issue in general textbooks on quantum mechanics
\cite{schiff81}. 
It is surely a basic concept of electron tunneling in nanoelectronics.
Actually the electron tunneling has been applied to scanning tunneling
microscopy, Josephson devices, superlattices, resonant tunneling
devices, and so on \cite{wolf85}. 
When we calculate the reflectivity and transmissivity, we solve
the time-independent Shr\"{o}dinger equation to obtain the 
wave functions of a stationary state, and then calculate the ratio of
the reflected and the transmitted probability current density to that
of incident flux. In these calculations, a boundary condition is not
varied with time.

 We are interested in the state where the boundary condition depends on
	time. 
This kind of problem is of interest for instance in expanding force
	fields \cite{berry84}, or in the evolution of metastable states in the
	early universe that is an interesting issue in cosmology
	\cite{lee04}. We treat in this article two problems of a finite or an infinite potential step moving with a constant velocity. These
	problems are the most basic concepts of quantum issues where boundary conditions are dependent on time.

\section{Infinite potential step}

\begin{figure}
\includegraphics{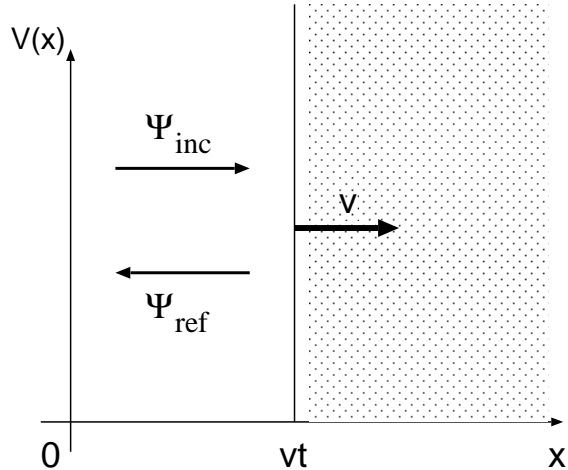}
\caption{\label{fig:pic_1} The schematic diagram of a infinite potential
 step.}
\end{figure}

	We solve the time-dependent Schr\"{o}dinger equation in one
	dimension as
\begin{equation}
 i \hbar \frac{\partial \Psi}{\partial t} = - \frac{\hbar^2}{2m} \frac{\partial^2 \Psi}{\partial x^2},
\label{eq-01}
\end{equation}
with a boundary condition that an infinite potential step is located at
$x = vt$ as shown in FIG. \ref{fig:pic_1}. It is because we can not use
the time-independent Schr\"{o}dinger equation with the boundary
conditions that depend on time. 
Although this analysis has been previously reported by Luan et
al. \cite{luan02}, we describe the essence of their theory here for the better
understanding of the finite potential analysis described in the next
section. We assume a general solution to be 
\begin{equation}
\Psi (x, t) = A e^{i(k_1 x - \omega_1 t)} +B e^{i(k_2 x - \omega_2 t)}, 
\label{eq-02}
\end{equation}
where the two terms correspond to an incident and a reflected wave
function, respectively. 
We should pay attention to $\frac{\hbar k_1}{m} \gg v $ that means the semi-classical point
of view. 
The solution (\ref{eq-02}) satisfies the Schr\"{o}dinger
equation (\ref{eq-01}) only when
\begin{equation}
\omega_1 = \frac{\hbar}{2m}k_1^2~~{\textrm{and}}~~\omega_2 = \frac{\hbar}{2m}k_2^2. 
\label{eq-03}
\end{equation} 

We consider two boundary conditions here. One condition is that the wave
function is zero (i.e. $A + B = 0$) at the boundary. The other condition
should be that the
first derivative of the wave function is also zero at the
boundary. Since the position of the boundary is a
function of $t$, the latter boundary condition can not be
used in the same manner as the boundary is fixed with time. We then give
an alternative boundary condition  that the phases of the incident and
the reflected wave functions at the boundary are the same i.e. $k_1 v -
w_1 = k_2 v - w_2$. This condition comes into
\begin{equation}
k_2 = -k_1 + \frac{2mv}{\hbar}, 
\label{eq-04}
\end{equation}
with the help of the equation  (\ref{eq-03}).
The expression  (\ref{eq-04}) can be well understood as 
a perfect elastic collision in a classical mechanics. From these
results, we can obtain the probability density:
\begin{equation}
|\Psi |^2 = 4 |A|^2 \sin^2 \left[\left( k_1 - \frac{mv}{\hbar}\right) \left( x-vt \right)\right]. 
\label{eq-05}
\end{equation}
On the other hand, the probability current density $j$ becomes
\begin{eqnarray}
j &=& \frac{\hbar}{2mi} \left( \Psi^* \frac{\partial \Psi }{\partial x} - \frac{\partial \Psi^*}{\partial x} \Psi \right) \nonumber \\ 
&=& 4 v |A|^2 \sin^2 \left(k_1 - \frac{mv}{\hbar}\right) \left(x - vt\right). 
\label{eq-06}
\end{eqnarray}
A group velocity can be calculated by dividing the probability current
density by the probability density. We find that the group velocity is equal to $v$.

\section{Finite potential step}

\begin{figure}
\includegraphics{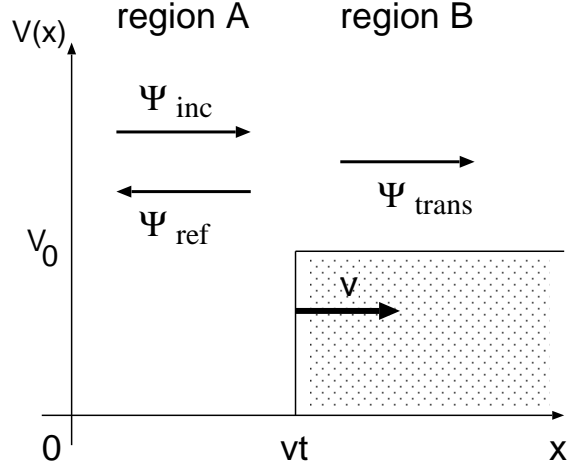}
\caption{\label{fig:pic_2} The schematic diagram of a finite potential step.}
\end{figure}

	We consider the other case that the potential step is as high as
	$V_0$ as
	shown in FIG. \ref{fig:pic_2}. The boundary is moving toward $+x$ with a speed of
	$v$. We should treat the two kinds of Schr\"{o}dinger equations in
	the two regions A and B as
\begin{equation}
i \hbar \frac{\partial \Psi_A}{\partial t} = - \frac{\hbar^2}{2m} \frac{\partial^2 \Psi_A}{\partial x^2}~~\left( x \leq vt:~\textrm{region~A} \right)\\
\label{eq-07}
\end{equation}
\begin{equation}
~~{\textrm{and}}~~i \hbar \frac{\partial \Psi_B}{\partial t} = - \frac{\hbar^2}{2m} \frac{\partial^2 \Psi_B}{\partial x^2}+ V_0 \Psi_B~~\left( vt < x:~\textrm{region~B} \right),
\label{eq-08}
\end{equation}
because transmitted wave function can exist in this case. We assume
general solutions in the two regions as
\begin{equation}
\Psi_A (x, t) = A e^{i(k_1 x - \omega_1 t)} +B e^{i(k_2 x - \omega_2 t)}
\label{eq-09}
\end{equation}
\begin{equation}
~~{\textrm{and}}~~\Psi_B (x, t) = C e^{i(k_3 x - \omega_3 t)}.
\label{eq-10}
\end{equation}
The first and the second terms in (\ref{eq-09}) correspond to an
incident and a reflected wave function, respectively. The solution
(\ref{eq-10}) corresponds to  the transmitted wave function. We can obtain 
\begin{equation}
\omega_1 = \frac{\hbar}{2m} k_1^2,~~\omega_2 = \frac{\hbar}{2m} k_2^2~~{\textrm{and}}~~\omega_3 = \frac{\hbar}{2m} k_3^2+\frac{V_0}{\hbar}
\label{eq-11}
\end{equation}
in the same way as the infinite potential step. We give two boundary
conditions as
\begin{equation}
A + B = C
 \label{eq-12}
\end{equation}
\begin{equation}
~~{\textrm{and}}~~k_1 v - \omega_1 = k_2 v - \omega_2 = k_3 v -\omega_3. 
\label{eq-13}
\end{equation}
The condition (\ref{eq-13}) derives from our assumption that the phase
of each wave function is the same. From the relationship of
(\ref{eq-13}) we can obtain the expressions:
\begin{equation}
k_2 = -k_1 + \frac{2mv}{\hbar}
 \label{eq-14}
\end{equation}
\begin{equation}
~~{\textrm{and}}~~k_3 = \frac{mv}{\hbar}+\sqrt{\left(k_1 - \frac{mv}{\hbar}\right)^2 - \frac{2mV_0}{\hbar^2 }}. 
\label{eq-15}
\end{equation}
When we substitute $v = 0$ in the expressions (\ref{eq-14}) and (\ref{eq-15}), we can
arrive at the well-known expressions for the potential step without
moving. The expression (\ref{eq-14}) describes the perfect elastic
reflection at the boundary similarly with the case of the infinite
potential step. In order to investigate the expression (\ref{eq-15}), we
should understand $k_1$ should be greater than $mv/\hbar$. This is required
for the collision of the incident wave function at the boundary. We
should pay attention to another critical point where the sign of the
expression inside the root in (\ref{eq-15}) is changed;
\begin{equation}
k_1 = \frac{mv}{\hbar}+ \frac{\sqrt{2mV_0}}{\hbar}. 
\label{eq-16}
\end{equation}
If the $k_1$ is greater than (\ref{eq-16}), the transmitted wave
function is oscillating, otherwise is of a damping oscillation. The
critical wave number is dependent on $v$. The first term in
(\ref{eq-16}) is the effect of the moving of the potential step. We can
understand that $\left( \sqrt{ \frac{m}{2}} v+\sqrt{V_0}\right)^2$ is the effective potential height of the step.

\subsection{ The case I ($k_1 > (\ref{eq-16})$)}

We consider the case that the transmitted wave function is
oscillating. Using the expressions (\ref{eq-11}), (\ref{eq-14}) and
(\ref{eq-15}), the probability density in the two regions are calculated
as
\begin{eqnarray}
|\Psi_A|^2 &=& |A|^2 +|B|^2 + A^* B e^{-i\left\{2\left( k_1 - \frac{mv}{\hbar} \right) \left( x -vt \right)\right\}} \nonumber \\
&+& A B^* e^{i\left\{2\left( k_1 - \frac{mv}{\hbar} \right) \left( x -vt \right)\right\}}
 \label{eq-17}
\end{eqnarray}
\begin{equation}
~~{\textrm{and}}~~|\Psi_B|^2 = |C|^2, 
\label{eq-18}
\end{equation}
where asterisks stand for the complex conjugate. On the other hand, the
probability current densities in the two regions become
\begin{eqnarray}
j_A &=& \frac{\hbar}{m} \left[ \left( k_1 |A|^2 + k_2 |B|^2 \right)  \right. 
\nonumber  \\
&+& \left. \frac{mv}{\hbar} \left\{ A B^* e^{i \left\{ 2 \left( k_1 - \frac{mv}{\hbar}\right) \left( x - vt\right)  \right\} } \right. \right. \nonumber \\
&+& \left. \left. A^* B  e^{-i \left\{ 2 \left( k_1 - \frac{mv}{\hbar}\right) \left( x - vt\right)  \right\} } \right\}  \right]
 \label{eq-19}
\end{eqnarray}
\begin{equation}
~~{\textrm{and}}~~j_B = \frac{\hbar}{m}k_3 |C|^2. 
\label{eq-20}
\end{equation}
The expression  (\ref{eq-19}) describes the sum of the incident and reflected probability current densities and the  (\ref{eq-20}) the transmitted one.
The complex coefficients of $A$, $B$, and $C$ can be generally expressed in the form
of
\begin{equation}
A = a e^{i \theta_a},~~B = b e^{i \theta_b}~~{\textrm{and}}~~C = c e^{i \theta_c}, 
\label{eq-21}
\end{equation}
respectively, where all variables are real values. We can describe the
boundary condition of (\ref{eq-12}) in the other form as
\begin{equation}
a + b = c~~{\textrm{and}}~~\theta_a = \theta_b= \theta_c. 
\label{eq-22}
\end{equation}
Since $|\Psi_A|^2$ and $|\Psi_B|^2$ are continuous at $x = vt$, we
obtain $(a+b)^2=c^2$ . On the other hand, the continuity condition of the probability current
density gives the relationship of
\begin{equation}
k_1 a^2 + k_2 b^2 + \frac{2mv}{\hbar} ab = k_3 c^2, 
\label{eq-23}
\end{equation}
and we find the two expressions:
\begin{equation}
\frac{b}{a} = \frac{k_1-k_3}{k_1+k_3-\frac{2mv}{\hbar}}
 \label{eq-24}
\end{equation}
\begin{equation}
~~{\textrm{and}}~~\frac{c}{a} = \frac{2k_1 - \frac{2mv}{\hbar}}{k_1+k_3-\frac{2mv}{\hbar}},
\label{eq-25}
\end{equation}
that arrive at well-known results when the potential step does not
move i.e. $v = 0$. 
By the way, the probability current density in the region A (the expression
(\ref{eq-19})) can be separated in the two components of the incident
and the reflected current densities as follows:
\begin{equation}
j_{inc} = \frac{\hbar}{m} k_1 |A|^2
 \label{eq-26}
\end{equation}
\begin{eqnarray}
~~{\textrm{and}}~~j_{ref} &=&- \frac{\hbar}{m} \left[ k_2 |B|^2   \right. 
\nonumber  \\
&+& \left. \frac{mv}{\hbar} \left\{ A B^* e^{i \left\{ 2 \left( k_1 - \frac{mv}{\hbar}\right) \left( x - vt\right)  \right\} } \right. \right. \nonumber \\
&+& \left. \left. A^* B  e^{-i \left\{ 2 \left( k_1 - \frac{mv}{\hbar}\right) \left( x - vt\right)  \right\} } \right\}  \right].
\label{eq-27}
\end{eqnarray}
We finally obtain reflectivity and transmissivity by dividing $J_{ref}$
and $J_B$ by $J_{inc}$:
\begin{equation}
R=\frac{-k_2 b^2 - \frac{2mv}{\hbar} ab \cos \left\{ 2 \left( k_1 - \frac{mv}{\hbar} \right) \left( x -vt \right)\right\}}{k_1 a^2}
 \label{eq-28}
\end{equation}
\begin{equation}
~~{\textrm{and}}~~T=\frac{k_3 c^2}{k_1 a^2}.
\label{eq-29}
\end{equation}
Using the expressions (\ref{eq-28}) and  (\ref{eq-29}), $R + T$ can be confirmed to be unity at
the boundary. 

	We can easily verify our results by considering the case of
	$v = 0$. The expressions (\ref{eq-24}) and (\ref{eq-25}) arrive at well-known results
	when $v = 0$. The expression of $T$ is the same as the one where
	$v = 0$. It is interesting that $R$
	depends on $x$ and $t$. The fact derives from the interference
	effect of the incident and reflected wave functions. 
The interference can occur only when the potential step is moving.
It is caused by the difference between the absolute values of $k_1$ and $k_2$.
We suppose
	that the effect can be applied to quantum wave interference
	devices. 

\subsection{The case II ($k_1 < (\ref{eq-16})$)}

	We consider the case that $k_1$ is smaller than the critical wave number
	of (\ref{eq-16}). We pointed out that transmitted wave function
	should be of a damping oscillation scheme. The $k_1$ should be
	larger than  $mv/\hbar$ in order that the group velocity of the
	incident wave function is larger than $v$ to reach the
	boundary. We assume that $k_1$ is far large than
	$mv/\hbar$ from the semi-classical point of view. The different point from the case I is that the $k_3$
	becomes an complex number as
\begin{eqnarray}
&& k_3 =  \gamma + i \beta, ~~{\textrm{where}}~~\gamma = \frac{mv}{\hbar}\nonumber \\
&& ~~{\textrm{and}}~~\beta=\sqrt{\frac{2mV_0}{\hbar^2}-\left(k_1-\frac{mv}{\hbar} \right)^2} > 0, 
\label{eq-30}
\end{eqnarray}
and then we obtain $\omega_3$ using the relationship  (\ref{eq-11}),
\begin{eqnarray}
&& \omega_3 = \sigma + i v \beta, \nonumber \\
&& {\textrm{where}} ~~ \sigma = \frac{\hbar}{2m}\left( k_1^2- \frac{2mv}{\hbar} k_1 + \frac{2m^2 v^2}{\hbar^2} \right).
\label{eq-31}
\end{eqnarray}

By substituting  (\ref{eq-30}) and  (\ref{eq-31}) for the expression  (\ref{eq-10}), we obtain the wave function in the region B as
\begin{equation}
\Psi_B = C e^{-\beta \left(x - vt \right) +i\left( \gamma x - \sigma t\right)}.
\label{eq-32}
\end{equation}
Therefore, the probability density and the probability current density
in the region B are expressed as
\begin{equation}
 |\Psi_B|^2=|C|^2 e^{-2 \beta \left( x - vt \right)}
\label{eq-33}
\end{equation}
\begin{equation}
~~{\textrm{and}}~~j_B = \frac{\hbar}{m} |C|^2 \gamma e^{-2 \beta \left( x - vt \right)}
\label{eq-34}
\end{equation}
Dividing (\ref{eq-34}) by (\ref{eq-33}) shows us that the group velocity is $v$. On the basis of the same discussion of (\ref{eq-21}) and (\ref{eq-22}) before, we can obtain $(a+b)^2=c^2$ also in this case, and draw the relationship of
\begin{equation}
 k_1 a^2 + k_2 b^2 +\frac{2 mv}{\hbar} ab = \gamma c^2
\label{eq-35}
\end{equation}
in stead of (\ref{eq-23}). Using $a+b=c$, we get simple relations of
\begin{equation}
\frac{b}{a} = 1,~~\frac{c}{a} = 2.
\label{eq-36}
\end{equation}
We substitute (\ref{eq-36}) for (\ref{eq-28}) and (\ref{eq-29}) to arrive at
\begin{equation}
 R = 1 - 4 \frac{mv}{\hbar k_1} \cos^2 \left( k_1 - \frac{mv}{\hbar} \right) \left( x - vt \right)
\label{eq-37}
\end{equation}
\begin{equation}
 T = 4 \frac{mv}{\hbar k_1} e^{-2 \beta \left( x - vt \right)}.
\label{eq-38}
\end{equation}
We can confirm easily that $R+T = 1$ at the boundary and that $R = 1$
and $T = 0$ if $v = 0$.  The semi-classical condition denoted in the previous section plays an important role here. It ensures that $T$ is less than unity.

\section{Conclusion}

We investigated the characteristics of the wave function and probability current density in a system with the potential step moving toward $+x$ direction at a constant velocity $v$. Since the position of the boundary depends on time, we solve the time-dependent Schr\"{o}dinger equation in one dimension. 
We used our boundary condition that the phases of the wave functions at the boundary are the same instead of ordinary condition that the first derivative of the wave function is the same at the boundary. We found the relation between the wave numbers of the incident and the reflected wave functions. The absolute value of the wave number is changed when the wave function is reflected at the boundary.  When the potential step is finite, the wave function can be transmitted if the energy of the incident wave function is larger than the effective potential height that is depend on $v$.

\bibliography{MovingPotentialStep}

\end{document}